\documentclass[iop]{emulateapj}
\usepackage{mathtools}
\usepackage{amsmath}
\usepackage{graphicx}
\usepackage{color}%

\newcommand{\kappapar}{{\kappa_\|}}
\newcommand{\kappaperp}{{\kappa_\bot}}
\newcommand{\diff}{\mathrm{d}}

\shorttitle{Cosmic-ray acceleration at perpendicular shocks}
\shortauthors{Takamoto \& Kirk}

\begin{document}

\title{Rapid cosmic-ray acceleration at perpendicular shocks
in supernova remnants}
\author{Makoto Takamoto$^{1,2}$}
\email{mtakamoto@eps.s.u-tokyo.ac.jp}
\author{John G. Kirk$^1$}
\email{john.kirk@mpi-hd.mpg.de}
\affil{$^1$Max-Planck-Institut f\"ur Kernphysik, 
Postfach 103980, 69029 Heidelberg, Germany\\
$^2$Department of Earth and Planetary Science, 
The University of Tokyo, 7-3-1~Hongo, Bunkyo-ku, Tokyo~113-0033, Japan}

\begin{abstract}
    Perpendicular shocks are shown to be rapid particle
    accelerators that perform optimally when the ratio $u_{\rm s}$ of
    the shock speed to the particle speed roughly equals the ratio
    $1/\eta$ of the scattering rate to the gyro frequency. We use
    analytical methods and Monte-Carlo simulations to solve the
    kinetic equation that governs the anisotropy generated at these
    shocks, and find, for $\eta u_{\rm s}\approx1$, that the spectral
    index softens by unity and the acceleration time increases by a
    factor of two compared to the standard result of diffusive shock
    acceleration theory. These results provide a theoretical basis for
    the thirty-year-old conjecture that a supernova exploding into the
    wind of a Wolf-Rayet star may accelerate protons to an energy
    exceeding $10^{15}\,$eV. 
\end{abstract}


\section{Introduction}
The \lq\lq knee\rq\rq\ at $\sim5\times 10^{15}\,$eV in the energy
spectrum of cosmic rays arriving at Earth defines a 
characteristic energy that can be
used to constrain the physics of the acceleration and propagation of
these particles.  Ideally, a theory of cosmic-ray
acceleration would provide a natural explanation of this
feature. Diffusive shock acceleration (DSA) in supernova remnants
(SNR), together with the associated amplification of the ambient
magnetic field, is in many respects a convincing theory, but it
fails to deliver a
simple prediction of the energy of the knee. The maximum energy to
which it can operate lies somewhat below $10^{15}\,$eV, and is thought to
be governed by the level at which the nonlinear
process of magnetic field amplification saturates. It is not known how
higher energy particles could be produced by within this theory 
\citep[for a recent review, see][]{2014BrJPh..44..415B}

The basic problem was identified long ago
\citep{1983A&A...118..223L,1983A&A...125..249L,2005JPhG...31R..95H}: 
DSA is too slow, because particles that cross and recross a shock front 
are accelerated at a rate $t_{\rm acc}^{-1}$, 
that is second order in 
the small parameter 
$\epsilon=u_{\textrm{s}}/v$:
$t_{\rm acc}^{-1}\sim\epsilon^{2}\omega_{\rm g}\,$, where 
$u_{\textrm{s}}$ is the 
shock velocity, $v$ the particle speed, 
and $\omega_{\rm g}=Z|e|Bc/E$ is the (angular) gyro-frequency of
a cosmic ray of charge $Ze$ and energy $E$ in a
magnetic field $B$.
Given this timescale, the only way for a SNR to
accelerate particles up to the knee is by generating a magnetic field
at the shock front that is substantially stronger than that in the
surrounding medium. Magnetic field amplification in SNR is consistent
with the observation of thin X-ray filaments in SNR 
\citep{2005A&A...433..229V,2006A&A...453..387P}, and emerges from numerical
simulations of parallel shocks using both particle-in-cell and hybrid
codes 
\citep{2012MNRAS.419.2433R,2013PhRvL.111u5003M,2014ApJ...783...91C,2015Sci...347..974M}.  
However, it 
appears to saturate at a level that does not permit 
acceleration to energies above the knee \citep{2013MNRAS.431..415B}.

In a seminal paper, \citet{1987ApJ...313..842J} 
(see also \citet{1988MNRAS.233..257O})
noted that this
problem is characteristic of parallel shocks, at which particles cross
and and recross the shock by diffusing along field lines, which leads
to a relatively long cycle time compared to the gyro period, particularly if
the scattering rate is low. At a
perpendicular shock, on the other hand, the cycle time is close
to the gyro period. 
If DSA were to remain valid, this would give an 
acceleration rate $t_{\rm acc}^{-1}\sim\epsilon \omega_{\rm g}$, 
which is fast enough to achieve energies well above the knee
\citep{1993A&A...271..649B}. The existence of a characteristic energy
at $5\times10^{15}\,$eV could then plausibly be associated with a
change in the confinement properties of the SNR
\citep{2011MNRAS.415.1807D,2013ApJ...768...73M}, since the gyro radius
of a proton of this energy in the interstellar magnetic field is
comparable with the radius of a SNR entering its Sedov phase of
expansion.

Early attempts to resolve this question using a Monte-Carlo approach
for nonrelativistic shocks were forced to make strong simplifications.
\citet{1993ApJ...409..327B}, for example, used a guiding center
approximation that implicitly assumes the distribution is almost
independent of gyro-phase. In later work \citep{1995ApJ...453..873E}
this was lifted, and agreement with the predictions of DSA over a limited range
of shock obliquities was found. However, 
a large-angle scattering algorithm was used that
strongly affects the ability to resolve anisotropies at the shock.
 \citet{1991aame.conf..291T} presented a method
that takes full account of anisotropy and non-conservation of the
first adiabatic invariant (magnetic moment), but neglected the
transport of particles across field lines --- a crucial ingredient for
the treatment of perpendicular shocks.  \citet{2006A&A...454..687M}
investigated both the spectrum and acceleration timescale for highly
oblique shocks, and found that both quantities agreed with the DSA
predictions even for low scattering rates.  However, they also used a
guiding center approach, which can be justified only if the scattering
rate is high.

Later work has lifted these artificial restrictions, although
the regime of nonrelativistic shock speeds and low scattering rates
remains challenging.  
The softening of the accelerated particle spectrum at a 
perpendicular shocks of
speed above $u_{\rm s}=0.1c$ was investigated quantitatively by 
\citet{2012ApJ...745...63S}, who used Monte-Carlo
simulations. Using a finite-difference method to
solve the equations that result from an expansion of the distribution
function in spherical harmonics, 
\citet{2013MNRAS.431..415B} also noted this effect for 
perpendicular shocks of speed above $u_{\rm s}=0.03c$. 
In both cases, the results are in
qualitative agreement with those presented below.  

An alternative approach, pioneered by \citet{1986ApJ...306..710D}
and pursued, for example, by 
\citet{1993MNRAS.264..248O} and \citet{1996JGR...10111095G}
consists of specifying the turbulent field around a shock front
explicitly, and examining the statistical properties of a large number
of trajectories computed by numerically integrating the equations of
motion. Evidence of an enhanced acceleration 
rate at oblique and perpendicular shocks was indeed found using this 
technique. Its advantages are that it 
allows insight to be gained at the microscopic level
by studying individual trajectories, and 
has the potential to include the effects of anomalous transport
\citep{2010ApJ...716..671L}, which 
are not easily accessible in a Monte-Carlo approach.

Our main goal here is a quantitative resolution of these issues.  We adopt a
plausible microscopic model of the scattering process and solve for
the full angular dependence of the distribution function, using both a
Monte-Carlo approach and an analytic approximation 
obtained in the limit of small shock speed and low scattering rate.
Our main results consist of the the spectral index and the
acceleration rate as functions of the shock speed, and the scattering rate.

In Section~\ref{qpshocks} we introduce the idealized model that is
used to describe particle transport, and, in this context, discuss DSA
at oblique and perpendicular shocks in more detail. The analytic
approximation is presented in Section~\ref{approximate} and the
Monte-Carlo method described in Section~\ref{MC}.  The main results
are presented in Section~\ref{results}, and their implications for the
acceleration of cosmic rays are discussed in Section~\ref{discussion}.
Section~\ref{conclusions} summarizes our conclusions.

\section{The transport model}
\label{qpshocks}

We consider the idealized situation of a plane shock front propagating
at constant speed into a uniformly magnetized plasma. Cosmic rays are
energetic charged particles, whose gyro radius is assumed to be large
compared to the thickness of the shock front, which can, therefore, be
treated as a discontinuity in the plasma velocity. 
The assumptions of
constant, uniform fluid velocity and magnetic field are justified
because we concentrate on the highest energy particles, whose energy
density is much smaller than that of the bulk of the cosmic rays, which,
in turn, is at most comparable to the ram pressure of the plasma
flowing into the shock.

In general, the cosmic-ray
distribution function $f(t,\vec{x},\vec{p})$ is a function of time
$t$, position $\vec{x}$ and momentum $\vec{p}$.  It is usual to assume
that particles are continually deflected by magnetic fluctuations
whose effect can be modeled as isotropic diffusion in the direction
of motion, i.e., as diffusion on the sphere of the end-point of the
unit vector $\vec{p}/p$, in addition to gyrating about the ambient
magnetic field.  This 
can be described by the Fokker-Planck equation,
 \citep[e.g.,][]{2011MNRAS.418.1208B} which, in the case of a
homogeneous magnetic field $\vec{B}$ embedded in a background plasma
that is at rest, reduces to
\begin{align}
  \label{kinetic}
\MoveEqLeft 
  \frac{\partial f}{\partial t} + 
\vec{v} \cdot \vec{\nabla} f + \omega_{\rm g} \frac{\partial f}{\partial \phi} 
\nonumber\\
&= \frac{\nu_{\mathrm{coll}}}{2} 
\left[ \frac{1}{\sin \theta} \frac{\partial}{\partial \theta} 
   \left( \sin \theta \frac{\partial f}{\partial \theta} \right)
 + \frac{1}{\sin^2 \theta} \frac{\partial^2 f}{\partial \phi^2} \right]
 ,
\end{align}
where $\vec{v}=\vec{p}/E$ is the CR velocity, $\theta$ and $\phi$ are
the spherical polar coordinates in momentum space with axis along
$\vec{B}$, and $\nu_{\rm coll}$ is a measure of the amplitude of the
fluctuations, which is usually called the \lq\lq collision
  frequency\rq\rq. The assumption of {\em isotropic} diffusion in angle
implies that $\nu_{\rm coll}$ is independent of $\theta$ and
$\phi$. In the following, we consider a situation in which the shock
front is a discontinuity that separates two half-spaces (upstream and
downstream) in each of which equation~(\ref{kinetic}) holds. Cosmic
rays are assumed to cross this discontinuity without deflection, so
that Liouville's theorem can be used to relate the distribution
immediately upstream (at $t=t_+$, $\vec{x}=\vec{x}_+$) to that
immediately downstream (at $t=t_-$, $\vec{x}=\vec{x}_-$):
\begin{equation}
\label{liouville}
f\left(t_+,\vec{x}_+,\theta_+,\phi_+,p_+\right)
=
f\left(t_-,\vec{x}_-,\theta_-,\phi_-,p_-\right)
\end{equation}
where $p_\pm$, $\theta_\pm$, and $\phi_\pm$ are the upstream and
downstream momentum coordinates that label the same momentum vector
just as $t_\pm$ and $x_\pm$ label the same space-time point. These
quantities are connected by a Lorentz boost, including, in general, 
a rotation to
take account of the change in the direction of $\vec{B}$ across the
shock.

Equations~(\ref{kinetic}) and (\ref{liouville}) suffice to describe
the particle acceleration process: solving the kinetic equation
(\ref{kinetic}) in the presence of a moving boundary (the shock front)
yields the particle residence times and escape probabilities, and
connecting the upstream and downstream solutions using Liouville's
theorem (\ref{liouville}) implements the boost in the energy measured
in the local fluid frame which each particle experiences when it
crosses the shock front. It is interesting to note that a very similar
system was analyzed by \citet{1963AnAp...26..234S}, who, however,
restricted the scattering to changes in phase, and considered only
those particles whose trajectories are almost tangential to the
shock. This limited the range of validity of the treatment to $\eta
u_{\rm s}\gg1$ (in the notation used below), thereby delaying the
discovery of DSA by fifteen years.

\subsection{The diffusion approximation}
If the cosmic ray distribution is almost isotropic,
the angular dependence in equation~(\ref{kinetic}) can be 
eliminated by expanding the momentum dependence of $f$ in spherical
harmonics \citep{1971RvGSP...9...27J,2004JCoPh.194....1K}. 
DSA is based on solving the resulting spatial diffusion equation in the presence
of a shock; a particularly useful introduction is given by
\citet{1983RPPh...46..973D}
, and this section  
reproduces the relevant results in order to facilitate the subsequent discussion.
Many analytic solutions are available,  
of which two are of particular interest here.
The first is for a steady-state particle distribution, 
with no cosmic rays far upstream and
no source term above momentum $p_0$. 
In this case, the distribution downstream does not depend on position, and
is a power-law in momentum:
\begin{align}
\label{solutionspectrum}
f(p)&\propto H\left(p-p_0\right)p^{-s}
\\
s&= 3r/\left(r-1\right)
\end{align}
where $r$ is the compression ratio of the shock and 
$H(x)$ is the Heaviside (or \lq\lq step\rq\rq) function.  Importantly, this
solution is independent of $\vec{B}$, so that, within the diffusion
approximation, shocks of all obliquities, including exactly 
parallel and perpendicular shocks, produce the same spectral index.
The second solution gives the mean time $\left<t\right>$ taken for a particle
to be accelerated from momentum $p_0$ to momentum $p_1$ at an oblique shock
\citep[see][Eq~3.31]{1983RPPh...46..973D}:
\begin{equation}
  \label{eq:2.2}
\left<t\right> = \frac{3}{u_+-u_-} 
  \int_{p_0}^{p_1}\, \frac{\diff p}{p}
\left( \frac{\kappa_+}{u_+} + \frac{\kappa_-}{u_-} \right)
  , 
\end{equation}
Here, 
$u_+$ ($u_-$) is the component of the plasma velocity upstream
(downstream) along 
the shock normal, in a frame in which the shock is at rest
and 
$\kappa_\pm$ are the corresponding components of the diffusion tensor: 
\begin{equation}
  \label{eq:2.3}
  \kappa_\pm = \kappapar_\pm \cos^2 \Psi_\pm + \kappaperp_\pm \sin^2 \Psi_\pm 
  , 
\end{equation}
with $\Psi_\pm$ 
the angle between the magnetic field and the shock normal. 
In this notation, the shock velocity $u_{\rm s}=u_+$ and the compression ratio
$r=u_+/u_-$. 

Applying the diffusion approximation to equation (\ref{kinetic}), 
\citet{1987ApJ...313..842J} showed
that the diffusion coefficients 
parallel and perpendicular 
to the magnetic field are related to the collision frequency by
\begin{equation}
  \label{eq:2.4}
  \kappapar = \eta \frac{r_g v}{3}, \quad \kappaperp = \frac{\eta}{1 +
    \eta^2} \frac{r_g v}{3} ,
\end{equation}
where 
$r_{\mathrm{g}}=v/\omega_{\mathrm{g}}$
and the dimensionless parameter
\begin{equation}
\label{etadeff}
\eta=\omega_{\mathrm{g}}/\nu_{\mathrm{coll}}
\end{equation}
describes the degree to which the particles are magnetized.

If $\eta$ is chosen to be a constant (independent not only of $\theta$
and $\phi$, but also of $p$), 
eq~(\ref{eq:2.2}), results in a mean acceleration rate proportional to
$p^{-1}$.  Discussions of DSA conventionally adopt this scaling
together with the additional restriction $\eta_\pm\ge1$.  Although the
applicability of these assumptions is disputed even within the
framework of the diffusion approximation
\citep{2009APh....31..237S,2014ApJ...792..133F}, we nevertheless adopt
them here, since our discussion focuses on the validity of the diffusion
approximation itself.

The acceleration rate $t_{\mathrm{acc}}^{-1}$ for a relativistic particle,
assuming, for simplicity, that $\eta$ is the same in the upstream and 
downstream plasmas, and that $p\gg p_0$, is given by equation~(\ref{eq:2.2}):
\begin{align}
t_{\mathrm{acc}}^{-1}\equiv&\left<t\right>^{-1}
\nonumber\\
=&\omega_{\mathrm{g}}
\frac{u_+^2}{c^2}\frac{r-1}{\eta r}
\left[
\cos^2\Psi_++\frac{\sin^2\Psi_+}{1+\eta^2}
+
\nonumber\right. \\
&
\left. \frac{r B_+}{B_-}\left(
\cos^2\Psi_-+\frac{\sin^2\Psi_-}{1+\eta^2}
\right)\right]^{-1}
\enspace.
\label{meantime}
\end{align}
At a parallel shock ($\Psi_\pm=0$, $B_-=B_+$) equation~(\ref{meantime}),
together with the restriction $\eta>1$ gives an upper limit on the acceleration
rate:
\begin{eqnarray}
\label{meantimeparallel}
t_{\mathrm{acc}}^{-1} &<& t_{\mathrm{B}}^{-1}\\
&=&\omega_{\mathrm{g}}
\frac{u_-^2}{c^2}\frac{r-1}{r(r+1)}
\nonumber\\
&\sim&\mathrm{O}\left(\epsilon^2\right)\omega_{\mathrm{g}}
\nonumber
\end{eqnarray}
commonly referred to as the {\em Bohm limit}. 
On the other hand, at a perpendicular shock
($\Psi_\pm=\pi/2$, $B_-=rB_+$)
\begin{equation}
\label{meantimeperp}
t_{\mathrm{acc}}^{-1} = t_{\mathrm{B}}^{-1}\frac{1+\eta^2}{\eta}\frac{1+r}{2}
\end{equation}
and the acceleration rate rises linearly with $\eta$ when $\eta\gg1$
\citep{1987ApJ...313..842J}.
This behavior applies for all shocks 
with $\cos\Phi_-\lesssim 1/\eta$, which are sometimes called 
{\em quasi-perpendicular}, but, for simplicity, we restrict ourselves 
in the following to exactly perpendicular shocks. 

\section{Approximate analytic solutions}
\label{approximate}

Using 
\lq\lq
mixed\rq\rq\ coordinates,
in which $\vec{p}$ (and $\vec{v}$, now expressed in units of $c$) 
are measured in the fluid rest frame, but $\vec{x}$ and $t$
are coordinates in a frame in which the shock is at rest,
the transport equation (\ref{kinetic}) becomes:
\begin{align}
\MoveEqLeft
\left(
1-v_zu \right)
  \frac{\partial f}{\partial t} + 
\left(v_z-u\right) c  \frac{\partial f}{\partial z}=  
\nonumber\\
&\frac{\omega_{\textrm{g}}}{\Gamma}
\left\{
- \frac{\partial f}{\partial \phi} 
+ \frac{1}{2\eta} 
\left[ \frac{\partial}{\partial \mu} 
   \left(1-\mu^2\right)\frac{\partial f}{\partial \mu} 
 + \frac{1}{1-\mu^2} \frac{\partial^2 f}{\partial \phi^2} \right]
\right\}
  \label{kineticshocka}
\end{align}
Here, the shock normal is along the $z$-axis, and we express both the 
component $v_z$ of the particle speed in this direction 
(measured in the fluid rest frame) as well as the 
speed $u$ of the fluid (assumed to be directed along the shock normal)
in units of $c$. The flow Lorentz factor is
$\Gamma=1/\sqrt{1-u^2}$, and 
spatial variations in the plane of the shock are ignored,
i.e., $\partial f/\partial x=\partial f/\partial y=0$.
The magnetic field is in the $y$-$z$ plane, and 
$\mu=\cos\theta$, so that 
$v_z=v\sin\theta\sin\phi$ is a function of $\mu$ and $\phi$. For cosmic rays,
the particle velocity, $v$, is close to unity. Note 
that equation~(\ref{kineticshocka}) is valid only for perpendicular shocks; 
the corresponding equation for subluminal shocks is given by
\citet{1989MNRAS.239..995K}.

Solutions that are stationary in the shock rest frame 
can be found by separating the variables in either
the upstream or the downstream region:
\begin{equation}
\label{expansiona}
f(z,\vec{p})=
F(p)\sum_{i} a_i \textrm{e}^{\Lambda_i z\omega_{\textrm{g}}/\Gamma c} Q_i(\mu,\phi)
\end{equation}
where $F$ is an arbitrary function of $p$, and the eigenvalues 
$\Lambda_i$ and eigenfunctions $Q_i$ obey
\begin{align}
\MoveEqLeft
\Lambda_i\left(v_z-u\right)Q_i
=
\nonumber\\
&\left\lbrace
-\frac{\partial}{\partial\phi}+\frac{1}{2\eta}\left[
\frac{\partial}{\partial \mu}\left(1-\mu^2\right)
\frac{\partial}{\partial\mu}+ 
\frac{1}{1-\mu^2}\frac{\partial^2}{\partial\phi^2}\right]
\right\rbrace Q_i
\label{eigenfunctionsa}
\end{align}
together with boundary conditions on $Q_i$ that ensure regularity and
single-valuedness on the sphere. In general, equation
(\ref{eigenfunctionsa}) has an infinite number of both positive and
negative discrete eigenvalues (labeled with $i>0$ and $i<0$
respectively), in addition to the eigenvalue $\Lambda_0=0$ with
eigenfunction $Q_0=\textrm{constant}$. These govern the spatial 
dependence of the solution: the isotropic part is independent of $z$, whereas
the eigenfunctions with $i<0$ decay exponentially towards positive $z$ (i.e., upstream), and grow exponentially downstream on a length scale that decreases as 
$\left|i\right|$ increases. Following the
procedure used for relativistic shocks \citep{1987ApJ...315..425K} the
solutions in the upstream and downstream regions can be
matched at the shock to find the function $F(p)$, which, in the 
absence of a source term, is a
scale-free power law $F\propto p^{-s}$. The boundary condition far upstream (at
$z\rightarrow\infty$ for $u_+>0$) is enforced by restricting the
expansion to $i<0$; that far downstream is
imposed by requiring the projection onto downstream eigenfunctions
with $i>0$ to vanish, thus preventing divergence of the distribution
as $z\rightarrow-\infty$. 

In the case of parallel, relativistic shocks, the expansion (\ref{expansiona})
was found to converge rapidly, and an accurate result was 
obtained by retaining only a single eigenfunction
\citep{2000ApJ...542..235K}. 
The current problem is more complicated, since the eigenfunctions are
functions of gyro-phase $\phi$ as well as pitch angle $\theta$, and the 
eigenvalue problem changes from a single-parameter ($u_{\rm s}$) 
problem into a two-parameter ($u_{\rm s}$, $\eta$) problem. 
This makes an expansion to high order cumbersome. Also, 
the differential operator in (\ref{eigenfunctionsa}) is not self-adjoint,
so that the adjoint operator (obtained by changing the sign of the 
$\partial/\partial\phi$ term) must be used to find the function
needed for projection onto the \lq\lq forbidden\rq\rq\ downstream
eigenfunction. 
Nevertheless, the analogy is close, so that one
can again hope to find a reasonable approximation
by using only a single eigenfunction. Adopting this approach,
the power-law index $s$ for a shock moving at speed $u_+$
into the upstream fluid and at speed $u_-$ in the downstream fluid
is given 
implicitly by the equation 
\begin{align}
\MoveEqLeft
S\equiv
\iint\diff\mu_-\diff\phi_- 
\bar{Q}_-\left(\mu_-,\phi_-\right)
\left({v_z}_- - u_-\right)\times
\nonumber\\
& \left(p_+/p_-\right)^{-s}Q_+\left(\mu_+,\phi_+\right) 
\nonumber\\
&= 0
\label{sconditiona}
\end{align}
where the label $i=-1$ of the retained, 
leading eigenfunction has been omitted, and the
notation $\bar{Q}$ is used to denote the corresponding 
eigenfunction of the adjoint
equation.  The notation $\left(\mu_\pm,\phi_\pm\right)$ is used 
to indicate angles in the up and downstream fluid frames.
For particles that move rapidly with respect to the shock,
(\ref{sconditiona}) can be
expanded to first order in $u/v$, leading to a linear equation for
$s$, with solution:
\begin{align}
\MoveEqLeft
s\approx
\frac{1}{\left(u_+-u_-\right)}\times
\nonumber\\
&\frac{\iint\diff\mu_-\diff\phi_- 
\bar{Q}_-\left(\mu_-,\phi_-\right)
\left({v_z}_--u_-\right)Q_+\left(\mu_+,\phi_+\right) 
} 
{\iint\diff\mu_-\diff\phi_- 
\bar{Q}_-\left(\mu_-,\phi_-\right)
{v_z^2}_-Q_+\left(\mu_+,\phi_+\right) 
}
\label{slinear} 
\end{align} 
where, for simplicity, the relative speed of the downstream fluid with
respect to the upstream fluid is assumed to be $u_+-u_-$,
implying that they both flow along the shock normal, which is the case for  
a shock of high Alfv\'enic mach number \citep[see, for example,][]{1989MNRAS.239..995K}.

The retained eigenfunction with 
$i=-1$ is special in two respects. Firstly, since 
it falls off more slowly with distance from the shock than do 
the other eigenfunctions, it is the dominant contribution to the 
distribution function far upstream, and so can have no zeroes
in the range $0<\phi<2\pi$, $-1<\mu<1$. 
Secondly, in the limit $u\rightarrow0$, it merges with the eigenfunction
$i=0$, which is a well-known feature characteristic of the 
diffusion approximation
\citep{1980JMP....21..740F}. Taking this limit, it is straightforward to show
that both $\Lambda_{-1}$ and 
the anisotropic part of $Q_{-1}$ are of first order in $u$. 
The transformation of
the arguments of the eigenfunction $Q_+$ in equation~(\ref{slinear}) 
from $\mu_+,\phi_+$ to
$\mu_-,\phi_-$ acts on the anisotropic part of this function only.
In the case of highly relativistic particles, to which we restrict ourselves 
in the following, it
produces a modification of this term that is of first-order in 
$\left(u_+-u_-\right)$.
It follows 
that 
\begin{eqnarray}
Q_+\left(\mu_+,\phi_+\right) 
&=& Q_+\left(\mu_-,\phi_-\right)+\textrm{O}\left({u_+}^2\right)
\end{eqnarray}
and, using the orthogonality relations
\begin{eqnarray}
\label{orthog}
\iint\diff\mu_+\diff\phi_+ 
\left({v_z}_+-u_+\right) \bar{Q}_- &=& 0\\
\iint\diff\mu_-\diff\phi_- 
\left({v_z}_--u_-\right) Q_+ &=& 0
\end{eqnarray}
in equation (\ref{slinear}),
leads immediately to the standard DSA result (\ref{solutionspectrum}).

However, this analysis is based on the assumption that $u$ is the 
only small parameter in the problem, which breaks down when scattering is 
sufficiently weak. Assuming the ordering $u\sim1/\eta\sim\epsilon\ll1$,
standard perturbation techniques (for details see appendix~\ref{perturbation}) 
lead to an expression for $Q$ that is anisotropic at zeroth order:
\begin{eqnarray}
Q&=&a\left(\mu\right)\textrm{e}^{\Lambda v\sqrt{1-\mu^2}\cos\phi}
+\textrm{O}\left(\epsilon\right)
\label{zerothordereigenfunction}
\\
\noalign{\hbox{where}}
a\left(\mu\right)&=&\textrm{Ps}^0_0\left(\mu,-\Lambda^2/2\right)
\label{spheroidalfunction}
\end{eqnarray}
with $\textrm{Ps}^0_0$ the angular, oblate, spheroidal wave function
with $m=n=0$ 
\citep[in the notation of][chapter~30]{2011ConPh..52..497T}.
The eigenvalue $\Lambda$ is related to the spheroidal eigenvalue 
$\lambda^0_0$ by
\begin{eqnarray}
\lambda^0_{0}\left(-\Lambda^2/2\right)
&=&\Lambda\left(\Lambda+2\eta u\right)
\label{lambda1eq}
\end{eqnarray}
and is of zeroth order in $\epsilon$. 

Evaluation of the $\phi$ integrals in (\ref{slinear}) 
is straightforward. The integrals over $\mu$ 
do not appear to be possible
analytically, but are simple numerical quadratures. An expansion
in powers of $\eta u$ yields 
\begin{eqnarray}
s
&=&\frac{3r}{r-1}+
\frac{9\left(r+1\right)}{20 r (r-1)}
\eta^2{u_+}^2
+\textrm{O}\left(
\eta^4{u_+}^4\right)
\label{sseries}
\end{eqnarray}
indicating a significant softening of the spectrum for finite $\eta u_+$, 
as compared to the standard DSA result. In section~\ref{results}
we compare this result and the angular dependence of the 
eigenfunction defined by equations (\ref{zerothordereigenfunction})
and (\ref{spheroidalfunction}) with the results of the 
Monte-Carlo simulations described below.

\section{Monte-Carlo Simulations}
\label{MC}

To solve the fully relativistic kinetic equation
(\ref{kinetic}), we use two conceptually different, but closely related
Monte-Carlo methods, similar to those described by
\citet{2001MNRAS.328..393A}
and by 
\citet{2004APh....22..323E} and \citet{2012ApJ...745...63S}.

In the first, we rewrite (\ref{kinetic}) in the form of a
Fokker-Planck equation, and use a theorem due to
\citet{ito1944} 
to write down a
stochastic differential equation governing a family of effective
trajectories, whose statistical properties are those of the required
solution $f$ \citep[see][chapter~4]{1994hsmp.book.....G}. 
An explicit first-order discretization scheme is then
used to construct a large number of these trajectories. 
In the second,
we start from the Boltzmann equation describing a distribution of
particles that move in the unperturbed, uniform magnetic field, 
but are subject to random collisions, each of which causes a small 
but finite angular deflection.
In the limit of small and frequent deflections 
this equation can be reduced to (\ref{kinetic}),
as we demonstrate in Appendix~\ref{boltzmann}.
but we solve it for small finite deflections,
advancing the trajectory in between the scatterings
by numerical integration. Details are provided in
Appendices \ref{sde} and \ref{boltzmann}.
We have verified that the results presented in Section
\ref{results} do not depend on which algorithm is employed.

In each case, trajectories are initiated at the shock front. They then
perform an excursion that either returns it to the shock, or terminates
the trajectory when it crosses a boundary placed at a large, 
fixed (in units of the gyro-radius $r_g$) distance 
from the shock front in the downstream region. An excursion that
returns to the shock front initiates a subsequent excursion in the
other half-space, starting at the same space-time point 
with the same space-time coordinates and momentum
four vector. 

The anisotropy of the particle distribution at the shock front is
the characteristic feature of this problem. In addition,
we are interested in two properties of the solutions: the
time-asymptotic power-law index at momenta well above injection, and 
the mean time taken for acceleration to a given momentum.
These quantities can be used to determine whether or not
a SNR lives long enough to enable 
acceleration at quasi-perpendicular shocks to play an important role.
Each of them can be extracted from a simulation of a large number of
trajectories, all injected at momentum $p_0$
(assumed $\gg mc$) at time $t=0$, by
recording the sets of values $t_i$ and $\vec{p}_i$ at
each crossing of the shock front. 
These values, collectively called {\em events}, are 
labeled by the integer $i$. 

The distribution at the shock front as a function of the angles
$\theta$ and $\phi$, and the momentum $p$ (in each case integrated
over the other two variables) is obtained by setting up
logarithmically spaced bins, and adding to these a weighting factor
equal to the reciprocal of the relative velocity of the particle with
respect to the shock surface.  
  Particles are initiated at the shock with an angular distribution
  that is arbitrarily chosen to be isotropic in the hemisphere
  entering the upstream plasma. However, after a few shock
  crossings, the bias thereby introduced is lost, and the distribution
  settles down to one that is independent of the number of crossings,
  and can be compared to the analytical form found for the scale-free,
  power-law distribution.  Provided the average number of crossings
per trajectory is large (in the example presented below it is roughly 100) 
the effect of the bias is not noticeable. 
  After a few crossings, particles display a power spectrum which
  extends from slightly above $p_0$ 
up to a momentum determined by the position of the
  downstream boundary, or the time limit placed on the trajectory.
  Within this range, the time-asymptotic power-law index $s$ is found
  from a least-squares fit in log-log space.  To find the average
  acceleration time, the events are again binned in $p$, and, at the
  same time, a running average of $t_i$ is accumulated. In the results 
presented below, $s$ and $t_i$ were determined using the range
$0.1<\log_{10}\left(p/p_0\right)<1$.  

\section{Results}
\label{results}

We present the results of fully relativistic simulations of
perpendicular shocks for an upstream speed $u_{\rm s}$ ($\equiv u_+$)
ranging from 0.01 to 0.2 (in units of $c$), and a
turbulence parameter $\eta$ between $1$ and $100$, and compare them to
analytic approximations. 
A total of 50,000 trajectories
were simulated for each set of parameters.
  For convenience, the plasma is assumed to flow along the shock
  normal both upstream and downstream, and the compression ratio
  $r=u_+/u_-$ is chosen to equal $4$, which corresponds approximately
  to the value derived from the Rankine-Hugoniot conditions for an
  ideal gas of specific heat ratio 5/3, (although it is not 
relativistically exact for
  any physically motivated equation of state).  The magnetic field
  strengths $B_+$ and $B_-$, measured in the up and downstream rest
  frames, respectively, are related by $B_+/B_- = r \Gamma_{\rm s}
  \sqrt{1 - (u_{\rm s}/r)^2}$ \citep[see][eq
  (4)]{1989MNRAS.239..995K}. Itoh's method was used in the
  simulations presented here. 

\begin{figure}[h]
 \input{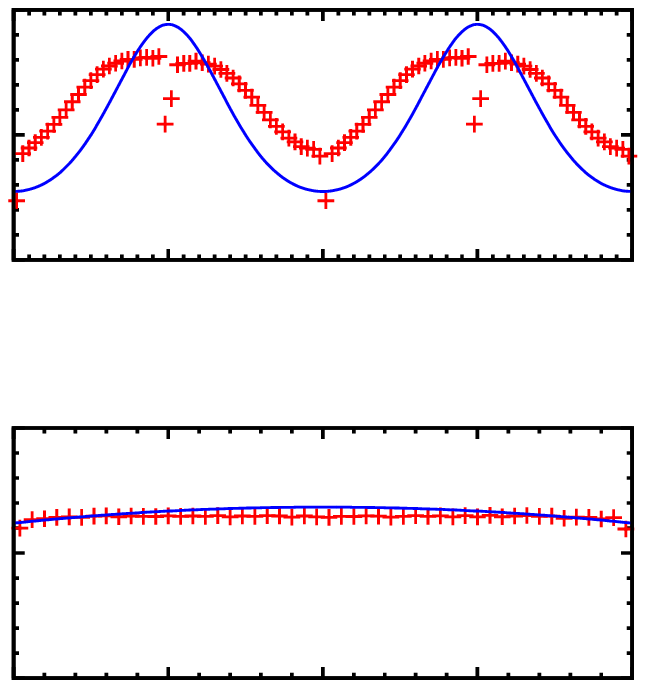} 
 \caption{The angular dependence of the distribution function $f$ at
   the shock (arbitrary normalization) for $u_{\textrm{s}}=0.012$ and
   $\eta=22$.  Monte-Carlo simulations (red points) are compared to
   zeroth-order analytic approximations. The top panel shows the phase
   dependence, ($f$ integrated over pitch angle), the bottom panel the
   pitch angle dependence, ($f$ integrated over phase). The blue lines are
   found by integrating the expression given in
   equation~(\ref{zerothordereigenfunction}) numerically. 
}
\label{angulardistribution}
\end{figure}

Figure \ref{angulardistribution} shows the angular distribution at the
shock front found from Monte-Carlo simulation as a function of
gyro-phase $\phi$ (top panel) and cosine $\mu$ of the pitch angle
(bottom panel), for upstream speed $u_{\textrm{s}}=0.012$ and turbulence
level $\eta=22$ ($\eta u_{\rm s}=0.26$). 
Particles are registered by the simulation as they cross
the shock front, so that very few events accumulate in bins where the
velocity vector lies close to the plane of the shock, i.e., at
$\sqrt{1-\mu^2}\sin\phi\approx u_{\textrm{s}}$. This accounts for the
relatively large fluctuations close to $\phi=0$, $\pm\pi$, 
which corresponds to 
grazing incidence for most values of $\mu$.
The
maximum of the distribution function lies close to 
$\phi=\pm\pi$, which corresponds to particles
moving along the shock front in the direction of the shock- or \lq\lq
grad-B\rq\rq- drift. The zeroth-order analytic approximation
reproduces the main features of the simulated distributions, although
it is more strongly peaked both in $\phi$. 
This may be either because the eigenfunction expansion 
requires several terms in order to converge to an accurate solution, and/or 
because the the analytic 
approximation is, strictly speaking, 
valid only in the limit of nonrelativistic shocks and 
low scattering rates ($u_{\rm s}\sim 1/\eta\rightarrow0$). 
In any case, it is clearly 
important to employ a scattering algorithm in the 
simulations that resolves angular structure on the small scale indicated
by the eigenfunction.

\begin{figure}[h]
 \input{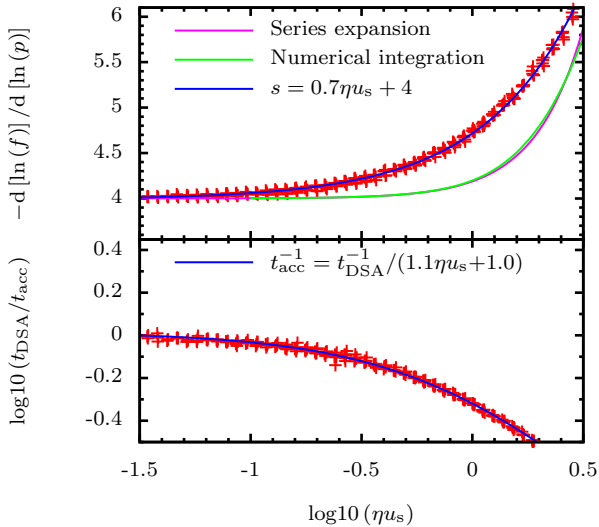} 
 \caption{The energy spectral index $s$ (top panel) and 
the acceleration rate divided by the DSA prediction (bottom panel),
as a functions of $\eta u_{\textrm{s}}$. Red crosses show
the results that fall in the plotted range
of Monte-Carlo simulations with 10 values of $\eta$, 
between 1 and 100, and 30 values of $u_{\textrm{s}}$ between 
0.01 and 0.2, in each case equally spaced in logarithms. 
Cyan and green curves in the top panel show analytic approximations
using the series expansion given in equation~(\ref{sseries}), 
and numerical integration respectively.
Blue curves show fits to the simulation results.
}
\label{powerlawrate}
\end{figure}

Figure \ref{powerlawrate} shows the time-asymptotic power-law index
$s$ (top panel) and the average acceleration rate (bottom panel) at a
perpendicular shock front, as functions of the product $\eta u_{\rm
  s}$, for various values of the parameters $u_{\rm s}$ and
$\eta$. In each case, the results fall close to a single curve 
with very small dispersion about it. This suggests
that the most important parameter in this range is the combination
$\eta u_{\rm s}$, which is also the only parameter of the approximate
solutions found in section \ref{approximate}. Physically, $\eta
u_{\rm s}$ is the collision time divided by the time
taken for the upstream flow to travel across one particle gyro-radius,
i.e., roughly the inverse of the number of collisions expected as an
undisturbed particle orbit advects across the shock. Provided the
fluid speed is nonrelativistic, this is the only physically
significant dimensionless quantity in the problem as we formulate it.

For $\eta u_{\textrm{s}}\ll 1$, the Monte-Carlo simulations show a
spectral index $s$ that lies close to the DSA prediction $s=4$, in
agreement with the analytic approximations. The acceleration rate in
this regime is also close to the DSA prediction, which is a factor
$2.5\eta\left(1+1/\eta^2\right)$ faster than the Bohm rate, as defined
in equation~(\ref{meantimeparallel}).  Inspection of the angular
distributions shows, as expected, that they are almost isotropic in
this region.

However, as $\eta u_{\textrm{s}}$ increases, simulations show that the
index $s$ increases (i.e.,~softens). This is also seen in the analytic
approximations, although the softening here sets in at somewhat higher
$\eta u_{\textrm{s}}$, and proceeds more rapidly.  According to the
simulation results, a softening of unity, i.e., $s=5$ is reached at
$\eta u_{\textrm{s}}\approx1$, whereas the zeroth-order analytic
approximation predicts this index at $\eta u_{\textrm{s}}\approx 2$.
The softening of the spectrum is accompanied by a reduction in the
acceleration rate, expressed in units of the DSA-predicted rate. At 
$\eta u_{\textrm{s}}=1$, the reduction is roughly a factor 2, implying that
acceleration still proceeds at a rate that is 
about a factor of $\eta$ faster than
the Bohm rate.

\section{Discussion}
\label{discussion}
The results presented above demonstrate that perpendicular shocks
impose a strong anisotropy on particles that are scattered in the
upstream and downstream plasma, when the ratio $1/\eta$ of the
scattering rate to the gyro frequency is less than or of the order of
the ratio of the shock speed to the particle speed. Instead of
diffusing in space around the shock front, as predicted by DSA, the
accelerated particles are then concentrated into a fan-beam structure
within a relatively small interval $\Delta\phi$ of gyro-phase directed
in the plane of the shock.  
This can be seen from the
  phase-dependence of the eigenfunction given in
  equation~(\ref{zerothordereigenfunction}): $Q \propto \exp[\Lambda v
  \sqrt{1 - \mu^2} \cos \phi]$.
The direction of the beam
corresponds to the drift imposed on an unscattered trajectory by the
presence of the shock front.  For $\eta u_{\rm s}>1$, the opening
angle of the fan beam can be estimated from the asymptotic expression
for the eigenvalue, 
  $\Lambda \sim - 3 \eta u_{\rm s}$ (see Appendix~\ref{perturbation}),
  to be $\Delta\phi\sim\left(\eta u_{\rm s}\right)^{-1/2}$. This
  roughly corresponds to the diffusive spread in phase produced by the
  scattering operator in equation~(\ref{kinetic}), when acting on a
  perfectly collimated beam over the time 
($\approx 1/\left(\omega_{\rm g}u_{\rm s}\right)$) taken for an unperturbed
  trajectory to cross the shock.

In this regime, a particle that is moving close to the shock front on its
downstream side which finds itself inside the fan beam has a relatively
high probability of recrossing the shock front many times. On the
other hand, a particle in the same position but with a momentum
directed out of the beam is likely to be swept away downstream after
only a few crossings. 
For this reason, the intuitive picture of DSA,
which is based on assigning all accelerated particles in the
downstream plasma an escape probability that is independent of their
position and direction of motion, is inadequate.  The more formal
derivation of DSA based on the diffusion-advection equation also
breaks down, because spatial diffusion describes the transport 
process only when the distribution is approximately isotropic. 

We find that the gradual breakdown of DSA at a nonrelativisic,
perpendicular shock as the collision rate decreases depends on the
single parameter $\eta u_{\rm s}$ and has two important effects on the
accelerated particles. Firstly, it softens the spectrum. For a
compression ratio of 4, the phase-space density, $f\propto p^{-s}$,
has $s\approx0.7\eta u_{\rm s}+4$, compared to the DSA prediction
of $s=4$. Secondly, it causes a reduction in the 
acceleration rate  by 
a factor of approximately $1.1\eta u_{\rm s}+1$ 
compared to the DSA prediction, which rises linearly with $\eta$.
Thus, optimal conditions for acceleration, in the sense that
it proceeds rapidly and produces a reasonably hard particle spectrum, 
are found when the anisotropies induced by the shock speed and the 
magnetic compression are comparable, i.e., when $\eta u_{\rm s}\approx1$.

The validity of DSA at perpendicular shocks has been a controversial
issue for many years.  \citet{1987ApJ...313..842J} suggested that
breakdown would happen when the collision time becomes longer than the
time during which a gyrating particle interacts with the shock,
leading to the approximate condition 
$\eta<1/u_{\rm s}$
and an upper limit on the acceleration rate that is first
order in $\epsilon$:
\begin{align}
t_{\mathrm{acc}}^{-1}
& 
< 1/\left(u_{\rm s}t_{\mathrm{B}}\right)
\nonumber\\
&=
u_{\rm s}\frac{r-1}{r}\omega_{\mathrm{g}}
\nonumber\\
&\sim\mathrm{O}\left(\epsilon\right)\omega_{\mathrm{g}}
.
\end{align}
This estimate is in rough agreement with our findings.
\citet{1994A&A...285..687A}, on the other hand, proposed that the collision 
frequency must be large enough to allow particles to diffuse along the magnetic
field whilst upstream of the shock. This leads to the restriction
$\eta<1/\sqrt{u_{\rm s}}$ and
\begin{align}
t_{\mathrm{acc}}^{-1}
 &<u_{\rm s}^{-1/2}t_{\mathrm{B}}^{-1}
\nonumber\\
&=
u_{\rm s}^{3/2}\frac{r-1}{r}
\omega_{\mathrm{g}}
\nonumber\\
&\sim\mathrm{O}\left(\epsilon^{3/2}\right)\omega_{\mathrm{g}}
.
\end{align}
  Our result that the acceleration time and spectral index are
  functions of the product $\eta u_{\rm s}$ does not support this
  conjecture. However, the situation may be different at oblique
  shocks, where particles have more opportunity to diffuse
  along the upstream magnetic field lines.  

The importance of these effects for the acceleration of high energy cosmic rays 
has been emphasized in particular by \citet{1987ApJ...313..842J}
and by \citet{1993A&A...271..649B}. Nevertheless, 
cosmic-ray acceleration at quasi-perpendicular shocks has received
relatively little attention for several reasons. Firstly, observations
of the polarization of radio emission in SNR~1006 show that
accelerated electrons are found predominantly in regions where the
magnetic field is disordered, and where the normal to the shock front
lies roughly in the direction of the external field, rather than
perpendicular to it \citep{2013AJ....145..104R}. This general trend is also
consistent with the predominantly radial orientation of the interior magnetic
field in young supernova remnants 
\citep{1993AJ....106..272R,2006A&A...453..387P,2012SSRv..166..231R}, and
fits in with the idea that magnetic field generation acts at
quasi-parallel shocks. 
Secondly, numerical simulations of 
acceleration using Monte-Carlo codes 
\citep{1995ApJ...453..873E,2004APh....22..323E,2012ApJ...745...63S}, 
hybrid codes \citep{2014ApJ...783...91C}, 
and PIC codes \citep{2012ApJ...755...68S,2013ApJ...771...54S,2013PhRvL.111u5003M}
also disfavor 
the quasi-perpendicular orientation, which, according to these
results, is less
efficient at injecting particles into the acceleration process,
provided the initial level of turbulence is very low.

However, these reasons do not directly apply to the problem of the
acceleration of very high energy ions, on which we focus our attention
in this paper.  On the one hand, observations of synchrotron radiation
relate exclusively to the electron distribution, which is likely to be
much more tightly confined to the shock front than are high-energy
ions.  On the other, numerical simulations consider an initially
uniform upstream magnetic field.  This may indeed inhibit injection,
but, in a more realistic situation, 
  some level of turbulence must be present initially. Our results
  indicate that efficient acceleration can be expected for an 
  effective collision frequency
  $\nu_{\rm coll}\sim u_{\rm s}\omega_{\rm g}$, which implies a very low 
  level of turbulence under SNR conditions. However, if the 
  collision frequency is even lower, we find that perpendicular shocks
  should produce only very steep spectra, in agreement with the
  apparent failure of these shocks to trigger acceleration in
  numerical simulations.
Furthermore, in a realistic situation,
low energy particles may see localized regions of
quasi-parallel geometry due to small length-scale fluctuations in the
upstream medium and/or the shock speed. These regions may be very
effective accelerators, possibly giving rise to substantial field
amplification. 
But, as described by \citet{2014BrJPh..44..415B} the
affected acceleration region remains limited in spatial extent, and
automatically provides an escaping flux of particles at the highest
energy to which it operates. In the context of the problem we consider
here, these can be regarded as being injected into an acceleration process
operating on larger spatial scales, on which the shock is
perpendicular.

Another 
  well-known 
argument against acceleration to energies above the
knee at perpendicular shocks is that particles drift across the shock
surface whilst undergoing acceleration, covering a distance
proportional to their energy. Depending on the geometry of the field,
this might move them out of the region where the shock is
perpendicular.  If one sets an upper limit to this distance equal to
the radius of the SNR, the corresponding upper limit on the energy
turns out to be comparable to that found for quasi-parallel shocks
  \citep[e.g.,][]{2014BrJPh..44..415B}, and is roughly equal to the
  energy a particle could gain by drifting from the pole to the equator 
(or vice versa) in
  a flow moving at the shock speed through a steady, magnetized,
  axisymmetric stellar wind.  

However, whether or not particle drift really limits the maximum
energy depends on the specific configuration of the magnetic field.
For example, in a uniform external field the drift motion induced by a
spherical shock front is directed along lines of constant latitude
(measured on the shock surface with the polar axis along the direction
of the external magnetic field), along which the shock does not change
its obliquity. Diffusion along the magnetic field lines in this
configuration might allow particles to escape to regions where the
shock is parallel, but this effect depends on the transport properties
in the upstream and downstream plasmas, rather than on the drift
motion.

The situation is different for a spherical shock front expanding into
a magnetic field that is anchored in the wind of a rotating
progenitor star. In the ideal case of a magnetic dipole aligned with
the rotation axis, a small region with quasi-parallel configuration
may exist close to the axis. However, if the progenitor is a
miss-aligned rotator, and/or one that undergoes repeated field
reversals, the undisturbed field is perpendicular to the shock normal
essentially everywhere, once this has expanded to well beyond the
Alfv\'en radius.  
In this case, the direction in which a particle
drifts as the shock moves over a magnetic field that is frozen into
the progenitor's wind, can be a rapidly changing function of the shock
radius.

The majority of supernovae are thought to result from the explosion of 
massive stars that can be assumed to have had a strong wind, and may have been
both magnetized and rapidly rotating. 
Assuming a wind speed of $v_{\rm w}$, and an Alfv\'en surface 
not too far from the star, 
the toroidal magnetic field at large radius $R$ 
can be estimated as $B(R ) = B_* \left(\Omega R_*/v_{\rm w}\right)
\left(R_*/R\right)$,  where 
$R_*$ is the stellar radius, $B_*$ the surface 
magnetic field and $\Omega$ the angular velocity of the star
\citep{1958ApJ...128..664P}. 
Then, defining the maximum energy
$E_{\rm max}$ to which a proton can be accelerated 
by identifying the acceleration rate 
given in equation~(\ref{meantimeperp}) with $u_{\rm s}c/R$,
setting $r=4$, and inserting values thought typical of Wolf-Rayet stars
\citep{2000A&A...357..283B},
leads to
\begin{align}
  \label{eq:pmax}
  E_{\rm max} &= \frac{3}{8}
\eta u_{\rm s}
\left(\frac{R_*\Omega}{v_{\rm w}}\right) eB_* R_*
\nonumber\\
&=1.7\,10^{16}\eta u_{\rm s}
\left(\frac{R_*\Omega}{v_{\rm w}}\right)
\left(\frac{B_*}{50\,\textrm{G}}\right)\left(\frac{R_*}{
3.10^{12}\,\textrm{cm}}\right)\, \textrm{eV}
  , 
\end{align}
We find that equation~(\ref{meantimeperp}) over-estimates the acceleration
rate by a factor of 2 when $\eta u_{\rm s}\approx1$, which is
unimportant in view of the uncertain
values of $\Omega$ and $B_*$. Therefore, we conclude that, 
under optimal conditions, i.e., when $\eta u_{\rm s}\approx1$, such objects 
may be capable of accelerating protons into a power-law spectrum somewhat 
softer than that predicted by DSA up to energies well in excess of 
$1\,\textrm{PeV}$.

\section{Conclusions}
\label{conclusions}
  We have studied the distribution function of particles accelerated
  at perpendicular shocks as their scattering rate decreases, finding
  that a strong anisotropy develops, which causes the theory of
  diffusive shock acceleration to fail. When the scattering time is
  comparable to the time taken for a particle orbit to traverse the
  shock front, the power-law index of the particle distribution is
  found to soften by roughly unity, compared to the DSA prediction and
  the acceleration rate is roughly halved.  These results provide a
  firm basis in kinetic theory for the long-standing conjecture that
  protons can be accelerated to energies well above $1\,\textrm{PeV}$
  at the perpendicular shocks that are expected to form when a
  supernova explodes into the wind of a massive progenitor.  

\acknowledgments 
We thank the anonymous referee for a very helpful report.  
This work is supported in part by the Postdoctoral Fellowships for Research 
Abroad program by the Japan Society for the Promotion of Science 
No.~20130253 and by the Research Fellowship for Young Scientists (PD) 
by the Japan Society for the Promotion of Science 
No.~20156571 (M.T.). 
M.T. also thanks the Theoretical Astrophysics group at the
Max-Planck-Institut fuer Kernphysik for their hospitality.

\appendix
\section{Approximate analytic solution} 
\label{perturbation}
In equation (\ref{eigenfunctionsa}) we assume 
$u\sim 1/\eta\sim\epsilon\ll1$, reintroduce the label $i$, and pose 
 an expansion:
\begin{eqnarray}
Q_i&\rightarrow& Q_i^{(0)}+\epsilon Q_i^{(1)}+\textrm{O}(\epsilon^2)\\
\Lambda_i&\rightarrow& \Lambda_i^{(0)}+\epsilon \Lambda_i^{(1)}
+\textrm{O}(\epsilon^2).
\end{eqnarray}
Since $v_z=v\sqrt{1-\mu^2}\sin\phi$, the zeroth order terms
\begin{equation}
\frac{\partial Q_i^{(0)}}{\partial\phi}+\Lambda_i^{(0)} v_z Q_i^{(0)}=0
\end{equation}
can be integrated to give (writing $\mu=\cos\theta$ where it makes the notation
more compact):
\begin{eqnarray}
Q_i^{(0)}&=&a_i\left(\mu\right)
\textrm{e}^{\Lambda_i^{(0)} \sin\theta\cos\phi}
\label{angulardist}
\end{eqnarray}
and, for the adjoint function
\begin{eqnarray}
\bar{Q}_i^{(0)}&=&\bar{a}_i\left(\mu\right)
\textrm{e}^{-\Lambda_i^{(0)} \sin\theta\cos\phi}.
\end{eqnarray}
The first order terms are:
\begin{eqnarray}
\left[\frac{\partial}{\partial\phi}
+\Lambda_i^{(0)} \sin\theta\sin\phi\right]Q_i^{(1)}
&=&\left(
\Lambda_i^{(0)}u-\Lambda_i^{(1)}\sin\theta\sin\phi \right)Q_i^{(0)}
+ 
\nonumber\\
&&
\frac{1}{2\eta}\left[
\frac{\partial}{\partial \mu}\left(1-\mu^2\right)\frac{\partial}{\partial\mu}+ 
\frac{1}{1-\mu^2}\frac{\partial^2}{\partial\phi^2}\right] Q_i^{(0)}.
\end{eqnarray}
Integrating, using the integrating factor 
$\textrm{e}^{-\Lambda_i^{(0)}\sin\theta\cos\phi}$ gives
\begin{eqnarray}
\label{q1eq}
Q^{(1)}_i&=&\textrm{e}^{\Lambda_i^{(0)} \sin\theta\cos\phi}
\left[\int_0^\phi\diff\phi'
F\left(\phi'\right)+ \frac{1}{4\eta}b\left(\mu\right)\right]\\
F\left(\phi\right)&=&
\textrm{e}^{-\Lambda_i \sin\theta\cos\phi}\left[
\left(\Lambda_i^{(0)}u-\Lambda_i^{(1)} \sin\theta\sin\phi\right)
Q_i^{(0)} + \phantom{\frac{1}{2\eta}}\right.
\nonumber\\
&&\left.\frac{1}{2\eta}\left(
\frac{\partial}{\partial \mu}\left(1-\mu^2\right)\frac{\partial}{\partial\mu}+ 
\frac{1}{1-\mu^2}\frac{\partial^2}{\partial\phi^2}\right) Q_i^{(0)}\right].
\end{eqnarray}
Imposing 
periodic boundary conditions in $\phi$, on $Q^{(1)}$ leads to 
\begin{equation}
\label{spheroidal}
\frac{\partial}{\partial \mu}\left(1-\mu^2\right)
\frac{\partial a_i}{\partial\mu}
+\frac{\Lambda_i^{(0)}}{2}\left(\Lambda_i^{(0)}\left(1+\mu^2\right)+4\eta u\right)a_i=0,
\end{equation}
which is to be solved with boundary conditions 
\begin{equation}
\frac{\diff a_i}{\diff\mu}= \mp
\frac{\Lambda_i^{(0)}}{2}\left(\Lambda_i^{(0)}+2\eta u\right)a_i\qquad\textrm{at\ }
\mu=\pm1.
\end{equation}
The first-order terms are then:
\begin{eqnarray}
Q_i^{(1)}&=&\frac{\textrm{e}^{\Lambda_i^{(0)}\sin\theta\cos\phi}}{4\eta}
\left\{b+4\eta\Lambda_i^{(1)}\sin\theta\left(\cos\phi-1\right)a_i -
\phantom{\frac{\diff a_i}{\diff\mu}}\right. 
\nonumber\\
&&\left.
\Lambda_i^{(0)}\sin\theta\sin\phi\left[\left(4+\Lambda_i^{(0)}\sin\theta\cos\phi\right)
a_i
+
4\cos\theta \frac{\diff a_i}{\diff\mu}\right]\right\}
\\
\bar{Q}_i^{(1)}&=&\frac{\textrm{e}^{-\Lambda_i^{(0)}\sin\theta\cos\phi}}{4\eta}
\left\{b-4\eta\Lambda_i^{(1)}\sin\theta\left(\cos\phi-1\right)a_i -
\phantom{\frac{\diff a_i}{\diff\mu}}\right. 
\nonumber\\
&&\left.
\Lambda_i^{(0)}\sin\theta\sin\phi\left[\left(4-\Lambda_i^{(0)}\sin\theta\cos\phi\right)
a_i
+
4\cos\theta \frac{\diff a_i}{\diff\mu}\right]\right\}
\end{eqnarray}
where $b\sim1$ and $\bar{b}\sim1$ are functions of $\mu$ which, together with 
$\Lambda_i^{(1)}$, can be 
constrained by examining the second order equations. However, because these
terms are even in $\phi$, they do not enter
into the computation of the index $s$.

Equation (\ref{spheroidal}) is a special case of the  
spheroidal differential
equation \citep[][chapter~30]{2011ConPh..52..497T}, whose solutions, for 
real $\Lambda_i$, are
the oblate, angular, spheroidal wave functions
$\textrm{Ps}_n^m\left(\mu,-\Lambda_i^2/2\right)$.
In general,
solutions exist for both positive and negative $\Lambda_i$, together
with the special isotropic solution $\Lambda_0=0$,
$a_0=\,$constant. However, to represent the upstream distribution we
require an eigenfunction which has no roots for
$-1<\mu<1$. 
This is the function
$\textrm{Ps}_0^0\left(\mu,-\Lambda_{-1}^2/2\right)$. 
The eigenvalues of the spheroidal wave equation 
$\lambda^m_n\left(\gamma^2\right)$ 
(in the notation of \citet{2011ConPh..52..497T})
are defined for $n\ge m$ and 
ordered such that $\lambda^m_m<\lambda^m_{m+1}<\dots$. 
They are related to $\Lambda_i$ by 
\begin{eqnarray}
\lambda^0_{-i-1}\left(-\Lambda_i^2/2\right)
&=&\Lambda_i\left(\Lambda_i+2\eta u\right)
\enspace.
\label{lambdaeq}
\end{eqnarray}
Thus, as expected, the largest negative eigenvalue
$\Lambda_{-1}$ corresponds to $\lambda^0_0$ and, therefore, to the 
eigenfunction that has no roots in the range $-1<\mu<1$.
For large $\eta\mu$, $\Lambda_{-1}\rightarrow -2\eta u$. 
A power series in $\eta u$, can be found directly from
(\ref{spheroidal}) 
\citep[see also][p~240]{meixnerschaefke54}:
\begin{eqnarray}
\Lambda&=&-3\eta u+\frac{3}{20}\left(\eta u\right)^3+ \textrm{O}\left(\eta^5u^5\right)
\\
a&=&1+\frac{3}{4}\mu^2\left(\eta u\right)^2 +
\frac{3}{160}\mu^2\left(4+9\mu^2\right)\left(\eta u \right)^4+ 
\textrm{O}\left(\eta^6u^6\right)
\end{eqnarray}
Using these series to evaluate the integrals in equation (\ref{slinear})
leads to equation (\ref{sseries}).

\section{Monte-Carlo method based on stochastic differential equations}
\label{sde}
Equation~(\ref{kinetic}), when rewritten in Fokker-Planck form, becomes:
\begin{equation}
  \label{fpform}
  \frac{\partial f}{\partial t} 
= 
-\frac{\partial}{\partial \vec{x}}\left(\vec{v}f\right) 
+\frac{\partial}{\partial \mu}\left(\frac{\mu f}{\eta}\right)
-\frac{\partial}{\partial \phi}f
+\frac{\partial^2}{\partial \mu^2}
\left(\frac{\left(1-\mu^2\right)f}{2\eta}\right)
+\frac{\partial^2}{\partial \phi^2}
\left(\frac{f}{2\eta\left(1-\mu^2\right)}\right)
\end{equation}
where 
 $\mu=\cos\theta$, 
time and space are measured in units of $\omega_{\mathrm{g}}^{-1}$ 
and $c/\omega_{\mathrm{g}}$, and $\vec{v}$ is in units of $c$..
Solutions to this equation can be found by constructing sample trajectories 
that satisfy the set of stochastic differential equations
\citep[see][]{1994hsmp.book.....G}
\begin{equation}
\begin{split}
\diff\vec{x}&= \vec{v}\diff t\\
\diff\mu&=-\left(\mu/\eta\right)\diff t 
+\left[\left(1-\mu^2\right)/\eta\right]^{1/2}\diff W_t\\
\diff\phi&=\diff t + \left[\left(1-\mu^2\right)\eta\right]^{-1/2}\diff W_t
\end{split}
\end{equation}
where $\diff W_t$ is an infinitesimal Wiener process.
We use an explicit first-order discretization to find a numerical solution
to this set, consisting, for each sample trajectory, 
of a sequence of points in phase space labeled by $i$:
\begin{equation}
\begin{split}
\vec{x}_{i+1}&= \vec{x}_i+\vec{v}_i\Delta t\\
\mu_{i+1}&=\mu_i-\left(\mu_i/\eta\right)\Delta t 
+\xi_i\sqrt{\frac{\Delta t\left(1-\mu_i^2\right)}{\eta}}\\
\phi_{i+1}&=\phi_i+\Delta t + \zeta_i\sqrt{\frac{\Delta t}{\left(1-\mu^2_i\right)\eta}}
\end{split}
\end{equation}
where $\xi_i$ and $\zeta_i$ are random numbers uniformly distributed
on the interval $\left(-\sqrt{3},\sqrt{3}\right)$,
which, therefore, have zero mean and unit variance. This scheme is
rapid, but has the disadvantage that trajectories that pass very close
to the points $\mu=\pm1$ are subject to errors. The affected range
depends on the time-step, and is given approximately by
$\left|\mu\right|> 1-\Delta t/\eta$. Typically, we choose $\Delta
t=10^{-2}$ and $1<\eta<20$, so that $\lesssim1\%$ of particles in an
isotropic distribution are affected. This is unimportant in the
simulations presented above, where no fine-scale structure in $\mu$ is
expected. However, it could become a concern for parallel,
relativistic shocks with $\Gamma\gtrsim100$.

Each sample trajectory starts on the shock front and consists of a
series of excursions into the up and downstream plasmas, ending when
it crosses a boundary placed a fixed distance $d$ downstream of the
shock.  We performed simulations for several different values of $d$.
For oblique shocks, $d\sim 200$ is sufficient, in the sense that 
a power-law distribution in $p$ is reproduced over several decades. However, 
parallel shocks
require $d\sim 1000$, reflecting the fact that the diffusion length
along the magnetic field is greater than that across it, if
$\eta>1$. When a timestep causes a trajectory to cross the shock
front, the step is repeated with smaller $\Delta t$ chosen to place 
the particle precisely on the shock front. The total number of timesteps
in the excursion is then checked, and the excursion repeated if this
number is too small --- typically less than 3. This procedure
eliminates trajectories that depend strongly on the finite size of the
timestep, at the expense of distorting the angular distribution of
particles at the shock front that move almost tangential to it, thereby 
limiting 
the ability of the simulation to resolve fine-structure in gyro-phase 
$\phi$ in directions close to the plane of the shock. 
From the analytic approximation, such structure should be present on the scale
$\Delta\phi\sim 1/\Lambda\sim \left(\eta u_{\rm s}\right)^{-1}$, so that the 
choice $\Delta t=1/100$ limits the accessible parameter range to 
$\eta u_{\rm s} < 100$.

\section{Monte-Carlo method based on the Boltzmann equation}
\label{boltzmann}

In this method
 we solve the equation that results when the right-hand
side of equation~(\ref{kinetic}) is replaced by the Boltzmann
collision operator corresponding to elastic pitch-angle scattering in
the test particle approximation, given by
\begin{equation}
  \label{eq:B11}
  \left( \frac{\partial f}{\partial t} \right)_{\rm coll} 
  = \frac{1}{t_{\rm scatt}} \left[ \frac{1}{2 \pi} \int^1_{-1} \int^{2 \pi}_0 
    d \mu' d \phi' f(\mu',\phi') p(\mu,\phi;\mu',\phi') 
                 - f(\mu,\phi) \right]
  ,
\end{equation}
where $t_{\rm scatt}$ is the mean time between scatterings 
and $p(\mu,\phi;\mu',\phi')$ is the probability that a particle
with $(\mu',\phi')$ is scattered into $(\mu,\phi)$ in a single 
scattering.
This equation is solved using the method
developed by 
\citet{1984ApJ...286..691E,1991SSRv...58..259J,2004APh....22..323E}.
Trajectories whose statistical
properties are given by $f$ are found by numerically integrating the
standard relativistic equations of motion in a steady background
electromagnetic field --- corresponding to the left-hand side of
equation~(\ref{kinetic}) --- using the exact solution or the Bulirsh-Stoer method over a 
fixed time interval chosen to equal the average time between 
scatterings: $t_{\rm scatt} = 2 \pi / \omega_g N$, where $N$ ($\gg1$) is a
parameter of the scattering model. (Formally, a random, exponentially 
distributed time interval with mean value $t_{\rm scatt}$ 
should be employed, but this is not necessary in the present problem, where
the number of scatterings between each encounter 
with the shock is required to be large). 
On scattering, trajectories are subject to a random
deflection through a small angle that is uniformly distributed between
zero and $\Theta_{\rm max}$ (for details see \citet{2012ApJ...745...63S}).
This enables one to identify a timescale 
$t_{\rm iso}$ on which the particle distribution would,
in the absence of a driving mechanism, relax to isotropy:
\begin{align}
t_{\rm iso}&=6t_{\rm scatt}/\Theta_{\rm max}^2
\end{align}
The equation of motion is solved in the upstream and downstream
half-spaces in the respective comoving frame, allowing the sample
trajectories to cross the shock front without deflection or energy
change.  The escape boundary is located in the downstream region
where the probability of particles returning to the shock front is
sufficiently small, as described in Appendix~\ref{sde}.

In order to demonstrate the link between this method and the
stochastic differential equation approach described in
appendix~\ref{sde}, we note that when the limit $N\rightarrow\infty$ 
is taken with $t_{\rm iso}$ fixed, the scattering model corresponds to
a phase function
\begin{align}
p\left(\mu,\phi;\mu',\phi'\right)&=p\left(\cos\Theta\right)
\nonumber\\
&= H\left(\cos \Theta_{\rm max}- \cos \Theta\right)/\left(
2\pi\Theta_{\rm max}\Theta
\right)  ,
\end{align}
where we assume $\Theta\ll1$ and use the normalization
$\iint d\phi d\Theta \sin\Theta p(\cos\Theta)=1$.
Expanding this phase function by writing
\begin{align}
  p(\cos \Theta) &= 4 \pi \sum^{\infty}_{n=0} \sum^{n}_{m=-n} \sigma_n 
  Y^m_n(\theta, \phi) Y^{m*}_n(\theta', \phi')
  , 
\label{eq:B13}
\end{align}
where $Y^m_n$ is a spherical harmonic, and the asterisk means the
complex conjugation, one finds
\begin{align}
  \sigma_n &= 
\frac{1}{2} \left[1 - \frac{\Theta_{\rm max}^2}{12}n (n+1) \right]
  .
  \label{eq:B17}  
\end{align}
Using equations (\ref{eq:B13}) and (\ref{eq:B17}), 
the collision operator, equation (\ref{eq:B11}), can be rewritten as follows: 
\begin{equation}
  \left ( \frac{\partial f}{\partial t} \right)_{\rm coll}
  = - \frac{\Theta_{\rm max}^2}{12t_{\rm scatt}} 
\int^{1}_{-1} \int^{2 \pi}_0 d \mu' d \phi' 
  \left[\sum^{\infty}_{n=0} \sum^{n}_{m=-n} n(n+1)
  Y^m_n(\theta, \phi) Y^{-m}_n(\theta', \phi')
  \right] f(\mu,\phi)
  \label{eq:B18}  
\end{equation}
where we used the completeness relation: 
\begin{equation}
  \label{eq:B19}
  \sum^{\infty}_{n=0} \sum^n_{-n} Y^m_n(\theta, \phi) Y^{-m}_n (\theta', \phi') 
  = \delta (\mu - \mu') \delta (\phi - \phi')
  . 
\end{equation}
Furthermore, 
using the following two relations: 
\begin{eqnarray}
   - n (n + 1) Y^m_n
  &=&
  \left[ \frac{\partial}{\partial \mu} (1 - \mu^2) \frac{\partial}{\partial \mu} 
  + \frac{1}{1-\mu^2}\frac{\partial^2}{\partial \phi^2} \right] Y^m_n 
  ,  
  \label{eq:B20}
  \\
  \int^{1}_{-1} \int_0^{2\pi}d \mu d\phi f(\mu) Y^m_n
  &=& 
  - \frac{1}{n (n + 1)} \int^{1}_{-1} \int^{2 \pi}_0 d \mu d \phi Y^m_n
  \nonumber 
  \\
  &\times& \left[ \frac{\partial}{\partial \mu} (1 - \mu^2) \frac{\partial}{\partial \mu} 
  + \frac{1}{1-\mu^2}\frac{\partial^2}{\partial \phi^2} \right] f 
  ,
  \label{eq:B21}  
\end{eqnarray}
the collision operator finally becomes: 
\begin{eqnarray}
  \label{eq:B22}
  \left ( \frac{\partial f}{\partial t} \right)_{\rm coll}
  &=& \frac{1}{2t_{\rm iso}} \left [\frac{\partial}{\partial \mu} (1 - \mu^2) 
    \frac{\partial}{\partial \mu}
    + \frac{1}{1 - \mu^2} \frac{\partial^2}{\partial \phi^2} \right] f(\theta, \phi)
,
\end{eqnarray}
which is equivalent to (\ref{kinetic}) one if we identify
$t_{\rm iso}=1/\nu_{\rm coll}$, or, equivalently,
$\eta=\omega_{\rm g}t_{\rm iso}$.
\\
\\


\end{document}